\documentclass[12pt]{article}

\def\kbras{{_k \langle }}
\def\dint{\int \hspace{-0,25cm} \int}
\def\3bras{{_3 \langle }}

\begin{document}

\title{On the Construction of Generalized Grassmann Coherent States}
  \author{}

\date{}

\maketitle

\vspace{4cm}

\center{\bf M. EL BAZ}\footnote{E-mail address: moreagl@yahoo.co.uk}

\center{\it Facult\'e des sciences, D\'epartement de Physique,
\linebreak Laboratoire de Physique Th\'eorique, LPT,
 \linebreak Av. Ibn Battouta, B.P. 1014, Agdal, Rabat, Morocco}

\vspace{5cm}

\abstract{A generalized definition of a deformation of the
fermionic oscillator ($k$-fermionic oscillators) is proposed. Two
prescriptions for the construction of generalized Grassmann
coherent states for this kind of oscillators are derived. The two
prescriptions differs in the nature of the generalized Grassmann
variables used. While we use Majid's definition for such variables
in the first case, Kerner's definition is used in the second one.}

\pagebreak

\section{Introduction}

The fermionic oscillator is generated by the operator of creation
and annihilation obeying the following anticommutation relations:
\begin{equation}\label{fermion}
\begin{array}{ccl}
\left\{ a, a^+\right\} &=& a\; a^+ + a^+ \; a = I \cr \left\{ N, a
\right\} &=& a \;\;\; ; \;\; \left\{ N , a^+\right\} = a^+
\;\;\;\;\; \hbox{with}\;\; N= a^+ \; a
\end{array}
\end{equation}
and a nilpotency condition:
\begin{equation}\label{pauli}
(a)^2 = (a^+)^2 =0
\end{equation}
 which is the
algebraic formulation of Pauli's exclusion principle for fermions.

At the algebraic level, a deformation of the harmonic oscillator
is an alteration of the (anti-)commutation relations between the
different operators. Generally it consists in introducing a set of
new dimensionless parameters, which by taking some special values
enable to recover the usual oscillator (fermionic or bosonic). In
this spirit, several deformations appeared in the literature,
either depending on one parameter \cite{arik, BiedMac, Greenberg}
or two parameters \cite{chakrabarti}.

However all this deformations have some common properties which made
it easy to unify all this deformations into a single generalized
deformation \cite{Daskaloyannis, Gazeau, Curado, 1}. This
unification is mainly done at the level of the Fock representation
and is totally encrypted into a function of the particle number. By
Choosing some special forms for this function one recovers a
particular deformation of the harmonic oscillator. A similar
generalization (unification) of the fermionic deformations
\cite{Chung} was proposed in \cite{bonatsos} where the deformation
is achieved using some function of the number operator with some
extra conditions compared to the bososnic deformations).

However this paper we are going to use a similar generalization to
that presented in \cite{1}, with some extra conditions again to
adapt it to a fermionic deformation. In fact a fermionic deformation
is also characterized by a degree of nilpotency of the different
operators\footnote{generalizing the nilpotency conditions
(\ref{pauli})}. It is worth mentioning that this nilpotency is
sometimes referred to as a generalized Pauli exclusion principle.
Which in Fock space representation reflects into the finite
dimensionality of this space.

It is because of this finite dimensionality that Grassmann variables
are needed in order to construct the relevant coherent states
\cite{klauder}.

It is known \cite{klauder, ohnuki} that in order to construct
fermionic coherent states one needs to use Grassmann variables to
take account of the nilpotency of the fermionic operators. It is
clear then that in order to construct coherent states related to
deformations of the fermionic oscillators and to generalizations of
these deformations one should use some generalized Grassmann
variables.

Kerner \cite{kerner} and Majid \cite{majid} Introduced such
variables that are characterized in one hand by non trivial
commutation relations and on the other hand by a degree of
nilpotency higher than the degree of nilpotency of the usual
Grassmann variables.

This paper is devoted to the construction of Generalized Grassmann
Coherent States GGCS, these are states related to a generalized
(unified) deformation of the fermionic oscillator by using the
generalized Grassmann variables. It is organized as follows:

In the second section, inspired from the results of \cite{1}, a
generalization of the fermionic deformations is proposed.

The third section is devoted to the construction of the Generalized
Grassmann Coherent States, first using Majid's definition of
generalized Grassmann variables (subsection 3.1) then using Kerner's
definition of the $Z_3$-graded Grassmann variables.

The different results are summarized and discussed in the last
section.

\section{Deformation of the Fermionic Oscillator}

A deformation of the fermionic oscillator, like a bosonic
deformation, consists in introducing some parameter(s) which for
some special values allow the recovering of the fermionic
oscillator. In addition to this it is characterized by a nilpotency
degree, $k\ge 2$, of the creation and annihilation operators. This
last property generalizes Pauli's exclusion principles by
authorizing multiparticle states.

After a thorough study of the different ways to unify deformations
of the bosonic oscillator \cite{Daskaloyannis, Gazeau, Curado, 1},
we find that the one which may suit a unification of deformations
of the fermionic oscillator is the one presented in \cite{1}.
Namely, a generalized deformed fermionic algebra, $ {\cal{F}}_q$,
is freely generated by operators of annihilation, creation and
identity $\{a, a^+, I\}$ with "q"-commutation relations:
\begin{equation}\label{qalgebra}
\begin{array}{ccccl}
\left[ a, a^+\right]_q &=& aa^+ - q \; a^+a &=& \Delta ' \cr
\left[a, \Delta\right]_q &=& a \; \Delta - q\; \Delta \; a &=&
\Delta' \; a \cr \left[ \Delta , a^+ \right]_q &=& \Delta \; a^+
- q \; a^+ \; \Delta &=& a^+ \; \Delta' \cr & \vdots&
\end{array}
\end{equation}
where $\Delta = a^+ \; a$ and $q$ is a complex parameter.
$\Delta'$ is generally interpreted as a $q$-derivative of $\Delta$
and is commuting with it.

Unlike a bosonic oscillator, a deformed fermionic oscillator is
characterized (in addition to the commutation relations) by a
degree of nilpotency of the operators of creation and
annihilation:

\begin{equation}\label{nilpotency}
(a)^k = \left(a^+\right)^k = 0\; .
\end{equation}
Generally, the degree of nilpotency $k$ depends on the value of
the parameter(s) of deformation.

It is easy to check at this stage that one can recover the
ordinary (non-deformed) fermionic oscillator (\ref{fermion},
\ref{pauli}) from the algebra above by choosing $q=-1$, $\Delta =
N$ and $\Delta'= I$ in (\ref{qalgebra}) and $k=2$ in
(\ref{nilpotency}). However, in order to obtain the different
deformations of the fermionic oscillator, it is convenient to work
in the Fock space representation of the algebra $ {\cal{F}}_q$. In
the following we will construct this representation space.

If we were to consider only (\ref{qalgebra}), then the Fock
representation would be the one given in \cite{1}; i.e., the Fock
basis vectors $\{ |n \rangle\}, \; \; n= 0,1,2\cdots$,
eigenvectors of the number operator, are constructed from the
vacuum state, $|0\rangle$ ($a|0\rangle = 0$), by successive
actions of the creation operator:
\begin{equation}
|n\rangle = { (a^+)^n \over ( \rho_n)^{1/2}} |0 \rangle
\label{nfock}
\end{equation}
where we have introduced the function $\rho_n$ which is defined
through the action of the different operators on the basis
vectors:
\begin{equation}
\begin{array}{ccl}
a \; |n\rangle &=& (\rho_n)^{1/2} |n-1\rangle  \cr a^+ \;
|n\rangle &=& (\rho_{n+1})^{1/2} |n+1\rangle \cr \Delta \;
|n\rangle &=& \rho_n \; |n\rangle \cr \Delta' \; |n\rangle &=&
\left( \rho_{n+1} - q \; \rho_n \right) \; |n\rangle
\end{array} \label{qFock}
\end{equation}

It is clear that in order to take account of the nilpotency
conditions (\ref{nilpotency}) some further conditions should be
imposed on the function $\rho_n$. In fact, the highest number
state one should be allowed to construct is $|k-1\rangle$ this
means that the function $\rho_n$ should be defined such that:
\begin{equation}
\rho_{k} = 0,
\end{equation}
$k$ being the nilpotency degree of the fermionic operators
(\ref{nilpotency}).

In this way, we recover the main difference between bosonic
oscillators (deformed or non-deformed) and fermionic ones: Pauli's
exclusion principle. Notice that, while at the algebra level the
generalized exclusion principle is implemented through the
nilpotency of the operators (\ref{nilpotency}), at the
representation level it is present through the finite dimensionality
of the Fock representation (this is in contrast to the infinite
dimension in the case of the bosonic deformation).

The algebra ${\cal F}_q$ defining a generalized fermionic
oscillator, one should be able to recover all the deformations of
the fermions oscillator as particular choices of the function
$\rho_n$. In the following we discuss some of the most famous cases.
\begin{itemize}
\item Ordinary Fermions: In this case $k=2$, and $q=-1$. Then choosing
$\rho_n= n \; , \; n= 0,1.$ one obtains the two-dimensional Fock
representation:
\begin{equation}
\begin{array}{ccccl}
a\; |0\rangle &=& 0 \;\; ; \;\;\;\; a\;|1\rangle &=& |0\rangle \cr
a^+\;|0\rangle &=&|1\rangle \;\; ; \;\;\;\; a^+ \; |1\rangle &=& 0
\cr
\end{array}
\end{equation}

\item Arik-Coon $q$-oscillator \cite{arik}:

This oscillator is defined by the following commutation relations:
\begin{equation}
[a,a^+]_q = aa^+ - qa^+a = I.
\end{equation}
When the complex parameter of deformation $q$ is a root of unity,
i.e. $q^k=1\; ;\;\; q= \exp{i{2\pi \over k}}$ ($k$ being an
integer) one obtains nilpotency of the operators $a$ and $a^+$ as
in (\ref{nilpotency}). This case is also a particular case of
${\cal F}_q$, as it is obtained from it by choosing
\begin{equation}
\rho_n = \{n\}_q = {1-q^n \over 1-q}
\end{equation}
and it is easy to check that $\rho_k = 0$.

\item Biedenharn-Macfarlane $q$-oscillator \cite{BiedMac}:
It is defined through the $q$-commutation relations:
\begin{equation}
aa^+ - q a^+a = q^{-N}.
\end{equation}
Here also when the parameter of deformation is a root of unity,
$q^k=1$ one obtains nilpotency of the operators $a$ and $a^+$. It
is recovered from ${\cal F}_q$ if one puts
\begin{equation}
\rho_n = [n]_q = {q^n - q^{-n} \over q - q^{-1}}
\end{equation}
and here also $\rho_k = [k]_q = 0$ because $q^k = 1$.

\item Chung and Parathasarathy \cite{chung}:

This deformation is based on the following $q$-deformed equation:
\begin{equation}
aa^+ + q' \; a^+a = q'^{N}\; ,
\end{equation}
it is recovered by putting $q'=-q$ in (\ref{qalgebra}) and
choosing:
\begin{equation}
\rho_n = q^{n-1} \, \hbox{sin}^2\left({n \pi \over k}\right) \; .
\end{equation}
One can check that, independently of the values of $q$,
$\rho_k=0$, which means that $a^k=(a^+)^k =0$.
\end{itemize}

\section{Generalized Grassmann Coherent States: GGCS}

\subsection{$1^{st}$ Case: using Majid's variables.}

According to Majid \cite{majid} a $Z_k$-graded Grassmann algebra
is generated by the variables obeying the following relations:
\begin{equation}
\begin{array}{cclc}
\theta_i \theta_j &=& q \; \theta_j \theta_i \; \;\;\;\; & \cr &&&
i,j= 1,2,\ldots  \;  \; \; i<j \cr (\theta_i)^k &=& 0
\end{array} \label{theta}
\end{equation}
where $q$ is a $k^{th}$ root of unity, i.e. $ \displaystyle q=
\hbox{e}^{2 \pi i  \over k}$:
\begin{equation}
q^k =1 \; , \; \; \; \bar q = q^{-1} = q^{k-1} \; , \;\;\; 1+ q+
\ldots q^{k-1} = 0 \; .
\end{equation}
One also introduces hermitian conjugate variables $\bar\theta =
\theta^{\dag}$:
\begin{equation}
\begin{array}{cclc}
\bar\theta_i \bar\theta_j &=& q \bar\theta_j \bar\theta_i \;
\;\;\;\; & \cr &&&  i<j \cr (\bar\theta_i)^k &=& 0\; .
\end{array}\label{bar}
\end{equation}
and we have the following commutation relations between the two
sectors:
\begin{equation}
\begin{array}{cclc}
\theta_i \bar\theta_j &=& \bar q \bar\theta_j \theta_i \; \;\;\;\;
& \cr &&& i<j \cr \bar\theta_i \theta_j &=& \bar q \theta_j
\bar\theta_i\, . \; \;\;\;\; & \cr
\end{array}\label{bartheta}
\end{equation}

All these relations can be written in a compact form as follows
\begin{equation}
\alpha_i \beta_j = q^{ab} \beta_j \alpha_i \;\;\;\; i<j
\label{majidgrassmann}
\end{equation}
where $\alpha$ and $\beta$ stand for $\theta$ and/or $\bar\theta$,
while $a$ and $b$ respectively denote the grades of $\alpha$ and
$\beta$; where to the $\theta's$ we attribute a grade 1 and a
grade $k-1$ to the $\bar\theta$'s.

In the following, and since we are dealing with a one mode
oscillator, we will be mainly interested in the one dimensional
case, which means that we will omit the subscripts.

We also have the following integration rules generalizing
Berezin's rules of integration:
\begin{equation}
\int d\alpha \; \alpha^n = \delta_{n,k-1} \; . \label{kBerezinint}
\end{equation}
where $\alpha = \theta  , \; \bar\theta$ and $ n$ is any positive
integer. And we have the following relations
\begin{equation}
\begin{array}{cclcrcl}
\theta d\bar\theta &=& q \; d\bar\theta \theta &
\;\;\;\;\;\;\;\;\;\;\;\; & \bar\theta d\theta &=& q \; d\theta
\bar\theta \cr \theta d\theta &=& \bar q \; d\theta \theta &
\;\;\;\;\;\;\;\;\;\;\;\; & \bar\theta d\bar\theta &=& \bar q \;
d\bar\theta \bar\theta \cr d\theta d\bar\theta &=& \bar q \;
d\bar\theta d\theta\, . && \cr
\end{array} \label{dtheta}
\end{equation}

These rules allow to compute the integral of any function  over
the Grassmann algebra written as a finite series in $\theta$ and
$\bar\theta$:
\begin{equation}
f(\theta, \bar\theta) = \sum_{i,j =0}^{k-1}C_{i,j} \theta^i
{\bar\theta}^j \, .
\end{equation}

Now in order to proceed further one need to define the behaviour
of these variables with respect to the oscillator operators i.e.
with respect to the creation and annihilation operators. In order
to do this one should be inspired by the (usual) fermionic case
where the Grassmann variables not only anticommute with each other
but anticommute also with the fermionic creation and annihilation
operators. Thus one should adopt general commutation relations
between the Generalized Grassmann variables and the Generalized
fermionic operators such that one recovers the anticommutativity
when $(k=2)$:
\begin{equation}
\begin{array}{rclcrcl}
\theta a^+ &=& q \; a^+ \theta & \;\;\;\;\;\;\;\;\;\;\;\; &\theta
a &=& \bar q \; a \theta \cr  \bar\theta a^+ &=& \bar q \; a^+
\bar\theta & \;\;\;\;\;\;\;\;\;\;\;\; &
 \bar\theta a &=& q \; a \bar\theta \; . \cr
\end{array} \label{atheta}
\end{equation}

Now we have all the ingredients to construct Generalized Coherent
States associated to the Generalized fermionic oscillator
${\cal{F}}_q$.

Coherent States (the canonical ones) have many properties and many
of these properties can be considered as defining properties for
general coherent states \cite{klauder}. And the defining property
for coherent states associated to oscillators is that they are
eigenstates of the annihilation operator with the eigenvalue given
by the label of the coherent states. So in our case we must find
states $|\theta \rangle_k$ such that
\begin{equation}
a |\theta \rangle_k = \theta |\theta \rangle_k \; .
\label{eigenstate}
\end{equation}
The states are written in a standard form in the Fock basis with
general coefficients:
\begin{equation}
|\theta \rangle_k = \sum_{n=0}^{k-1} \alpha_n \theta^n |n\rangle
\end{equation}
Now to find the states $|\theta \rangle_k$ one should find the
coefficients $\alpha_n$ for which the property (\ref{eigenstate})
is verified.
\begin{eqnarray}
a |\theta \rangle_k &=& \sum_{n=0}^{k-1} \alpha_n a \theta^n
|n\rangle \nonumber \\ &=& \sum_{n=0}^{k-1} \alpha_n q^n \theta^n
a |n\rangle \nonumber \\ &=& \sum_{n=1}^{k-1} \alpha_n  q^n
\theta^n \left(\rho_n\right)^{1\over 2} |n-1\rangle \nonumber \\
&=& \theta \sum_{n=0}^{k-2} \alpha_{n+1}  q^{n+1} \theta^n
\left(\rho_{n+1}\right)^{1\over 2} |n\rangle \nonumber
\end{eqnarray}
which by definition should equal$$  \theta |\theta \rangle_k =
\theta \sum_{n=0}^{k-1} \alpha_{n} \theta^n |n\rangle$$ So the
condition on $\alpha_n$ for the equation (\ref{eigenstate}) to be
verified is:
\begin{equation}
\alpha_n = \alpha_{n+1} \, q^{n+1} \,
\left(\rho_{n+1}\right)^{1\over 2} \; .
\end{equation}
Iteratively solving this equality yields the solution:
\begin{equation}
\alpha_n = {1\over \left(\rho_{n+1}\right)^{1\over 2}} {\bar
q}^{n(n+1)\over 2}
\end{equation}
Then the GGCS associated to $ {\cal{F}}_q$ are written in the Fock
basis as:
\begin{equation}
|\theta \rangle_k = \sum_{n=0}^{k-1} {\bar q}^{n(n+1) \over 2} \;
{\theta^n \over \big(\rho_n!\big)^{1\over 2}} \; |n\rangle \; .
\label{kCS2}
\end{equation}
Using equation (\ref{nfock}) one can see that the states $|\theta
\rangle_k$ are generated from the vacuum through:
\begin{equation}
|\theta \rangle_k = \sum_{n=0}^{k-1} {\bar q}^{n(n+1) \over 2} \;
{\theta^n (a^+)^n \over \big(\rho_n!\big)} \; |0\rangle \; .
\end{equation}
Using equation (\ref{atheta})one can prove that:
\begin{equation}
  \theta^n (a^+)^n = q^{{n(n+1) \over 2}} \,(a^+ \theta )^n
\end{equation}
which permit us to rewrite the GGCS in a more conventional way:
\begin{equation}
|\theta\rangle_k = \sum_{n=0}^{k-1} {(a^+ \theta)^n \over \rho_n!}
\; |0\rangle \; := \; \hbox{exp}_q(a^+ \theta) \; |0\rangle \; .
\label{kCS}
\end{equation}
where the generalized deformed exponential function introduced is
defined through:
\begin{equation}
\hbox{exp}_q(x) = \sum_{n=0} {x^n\over \rho_n!} \label{qexponential}
\end{equation}

Next we will prove that the GGCS, as all coherent states, do provide
a resolution of unity \cite{klauder}. We will look for a resolution
of unity in the form:
\begin{equation}
\dint d\bar\theta d\theta \; \omega (\bar\theta \theta) \;
|\theta\rangle_k{\kbras}\theta| = I \; , \label{kRI}
\end{equation}
where the weight function is written as follows
\begin{equation}
\omega(\bar\theta \theta) = \sum_{n=0}^{k-1} \; c_n \; \theta^n
{\bar\theta}^n \; . \label{weight1}
\end{equation}
Then the problem of proving the resolution of unity reduces to
finding the adequate weight function, that is, to finding the
coefficients $c_n$ such that (\ref{kRI}) holds:

By replacing (\ref{weight1}) and (\ref{kCS2}), the LHS of
(\ref{kRI}) can be written as follows:
\begin{equation}
\dint \sum_{l,n,p=0}^{k-1} d\bar\theta d\theta \; c_n \theta^n
{\bar\theta}^n \; {\bar q}^{l(l+1)\over 2} q^{p(p+1)\over 2} \;
{\theta^l {\bar\theta}^p \over \big(\rho_l!\big)^{1\over 2}
\big(\bar\rho_p!\big)^{1\over 2}} \; |l\rangle\langle p| \; .
\end{equation}
Taking account of the integration rules and the $q$-commutation
rules it becomes:
\begin{equation}
\dint d\bar\theta d\theta \; \sum_{l,n=0}^{k-1} {c_n q^{nl} \over
|\rho_l|!} \; \theta^{n+l} {\bar\theta}^{n+l} |l\rangle\langle
l|\; .
\end{equation}
which in turn by using the completeness of the Fock space basis
$\displaystyle \sum_{l=1}^{k-1} |l\rangle\langle l| = I$ becomes:
\begin{equation}
I= \sum_{\stackrel{n,l=0}  {n+l = k-1}}^{k-1} {c_n q^{nl} \over
|\rho_l|!} \dint d\bar\theta d\theta \theta^{n+l}
{\bar\theta}^{n+l} \; |l\rangle\langle l| \, .
\end{equation}
Since this should give the identity operator, one has the
following constraint on the coefficients $c_n$:
\begin{equation}
{c_n q^{nl} \over |\rho_l|!} = 1 \;\; \hbox{and} \;\;\; n+l = k-1
\;.
\end{equation}
i.e.
\begin{equation}
\begin{array}{ccl}
c_n &=& {\bar q}^{n(k-n-1)} |\rho_{k-n-1}|! \cr &=& q^{n(n+1)}
|\rho_{k-n-1}|! \cr
\end{array}
\end{equation}
The weight function appearing in the resolution of unity
(\ref{kRI}) is therefore given by
\begin{equation}
\omega(\bar\xi \xi) = \sum_{n=0}^{k-1} q^{n(n+1) \over 2}
|\rho_{k-n-1}|! (\bar\xi \xi)^n\; .
\end{equation}

\subsection{$1^{st}$ Case: using Kerner's variables.}

Kerner's starting point for generalizing the Grassmann variables is
somehow the same as Majid's. In fact, both generalizations start
from the fact that the Grassmann algebra is a $Z_2$-graded algebra.
Then suggesting that a generalized Grassmann algebra should be a
$Z_k$-graded one, where $k$ is a positive inee=ger $k\ge 2$. However
the procedures diverge in the way of implementing this $Z_k$-graded
structure on the constructed algebra. In fact, Majid implement this
by replacing -1 (the second root of unity) by a $q= \hbox{exp}{2pi i
\over k}$ (a $k^{th}$ root of unity). Thus changing the Grassmann
anticommutation relation by a $q$-commutation relation
(\ref{majidgrassmann}). While Kerner's point of view is
fundamentally different \cite{kerner}:

The $Z_2$-graded structure of the Grassmann algebra reflects itself
in the fact that $Z_2$ (the cyclic group) is a symmetry group for
this algebra; that is for any two elements of the Grassmann algebra
the following property holds
\begin{equation}
{\cal{Z}}_2(\xi_i\xi_j) = \xi_i\xi_j + \xi_j \xi_i  = 0\;\;\;\,\;\;
\hbox{where} \;\;\;\; {\cal{Z}}_2 \; \hbox{ is an element of } Z_2
\end{equation}
(from which, one deduces the anticomutativity property of the
Grassmann variables and their nilpotency).

A generalized $Z_k$-graded Grassmann algebra then should generalize
this property by imposing a symmetry with respect to the cyclic
group $Z_k$. Kerner considered (and totally solved) the case where
$k=3$. In this case the $Z_3$-graded Grassmann variables obey the
following ternary property (symmetry with respect to the cyclic
group $Z_3$):
\begin{equation}
{\cal{Z}}_3(\xi_i\xi_j\xi_k) = \xi_i\xi_j\xi_k + \xi_j\xi_k\xi_i +
\xi_k\xi_i\xi_j =0 \;\;\;\;\; \;\;\;\; {\cal{Z}}_3 \; \hbox{ being
an element of } Z_3 \label{z3symmetry}
\end{equation}

A particular solution of the constraint (\ref{z3symmetry}) is given
by the ternary relation:
\begin{equation}
\xi_i\xi_j\xi_k = j \; \xi_j\xi_k\xi_i = j^2 \; \xi_k\xi_i\xi_j
\label{xikerner}
\end{equation}
i.e. under a cyclic permutation of the elements of a ternary product
a factor $j$ appears where $j$ is a cubic root of unity:
\begin{equation}
j = \hbox{exp}\left({2\pi i \over 3}\right) \;\; ; \;\; j^3=1 \;\; ;
\;\; \bar{j}=j^2 \;\; ; \;\; j^2+j+1=0
\end{equation}
So the $Z_3$-graded Grassmann variables is obtained by assuming that
there are no binary relations among these variables (i.e. products
of the form $\xi_1 \xi_2$ and $\xi_2 \xi_1$ are considered as
independent elements). Instead of this, ternary relations
(\ref{xikerner}) are given.

Moreover, two important properties follow automatically from
(\ref{xikerner}):
\begin{itemize}
\item $\displaystyle \left( \xi_i\right)^3 = 0$

\item $\displaystyle \xi_i\xi_j\xi_k\xi_l = 0$
\end{itemize}

One also introduces grade-2 elements \cite{kerner}, which are duals
to the $\xi$'s and obey similar relations with $j$ replaced by
$j^2$:
\begin{equation}
\bar\xi_i \bar\xi_j \bar\xi_k = q^2 \; \bar\xi_j \bar\xi_k \bar\xi_i
= q \; \bar\xi_k \bar\xi_i \bar\xi_j
\end{equation}

There exist binary relations but only for a product involving a
grade-1 element and a grade-2 one:
\begin{equation}
\xi_i \bar\xi_j = q\; \bar\xi_j \xi_i \label{barxi}
\end{equation}

In the following, and as in the previous section, we will omit the
subscripts as we will be dealing mainly with the one mode oscillator
and a one dimensional Grassmann algebra.

Integration over this algebra is carried with the same relations
(\ref{kBerezinint}). In this case ($k=3$), these are written
explicitly:
\begin{eqnarray}
\int d\xi . 1  = \int d \bar{\xi} . 1 &=& \int  d \xi \,.\, \xi = \int d \bar{\xi } \,.\,  \bar{\xi}  = 0 \\
\int d \xi \, . \,\xi^2 &=& \int d\bar {\xi} \,.\, \bar {\xi}^2 = 1
\nonumber
\end{eqnarray}
and here also, the rules allows to compute the integral of any
function over the Grassmann algebra.

In order to proceed further one need to define the behavior of these
variables with respect to the generalized fermionic (creation and
annihilation) operators, i.e. relations similar to (\ref{atheta}).
To write these relations, one should first note that there is a deep
difference between the two generalized Grassmann algebras discussed.
In fact, grading is the main rule governing the relations in
Kerner's definition of the generalized Grassmann variables. So in
order to write relations similar to (\ref{atheta}) in this case one
has to introduce a grading over the generalized fermionic oscillator
algebra (\ref{qalgebra}, \ref{nilpotency}). A way of doing this, is
to impose that the coherent states to be constructed should be
grade-0 \cite{1, chung}. Then the grading that follows from this
requirement is that $a$ is a grade-1 while $a^+$ is a grade-2
element.

Combining this fact with what was announced previously leads to the
conclusion that no binary relations can be imposed on $\xi$ and $a$
(or $\bar\xi$ and $a^+$), because no binary relations exists between
elements with the same grade. These relations are however allowed
between elements with different gradings:
\begin{equation}
\xi \, a^+ = j \; a^+ \, \xi \;\;\; , \;\;\; \bar\xi \, a = j^2 \; a
\, \bar\xi \label{axi}
\end{equation}

It is worth mentioning that since we restricted the definition of
the generalized Grassmann variables to the $Z_3$-graded case
($\xi^3=0$), the coherent states we shall construct are associated
to the generalized fermionic oscillator (\ref{qalgebra},
\ref{nilpotency}) with $k=3$ i.e. ($a^3=(a^+)^3 = 0$); the
nilpotency of the generalized Grassman variables is intimately
related to the nilpotency of the fermionic operators, i.e. to the
generalized Pauli's exclusion principle.

\vspace{0.5cm}

{\underline{$Z_3$-graded Coherent States}}

\vspace{0.5cm}

The coherent states to be constructed should be constructed as
eigenstates of the annihilation operator:
\begin{equation}
a|\xi \rangle_3 = \xi \; |\xi \rangle_3 \; . \label{3eigenstates}
\end{equation}
Proceeding as in the previous section, by first writing
$|\xi\rangle_3$ in its general form in the representation space of
the generalized fermionic oscillator algebra\footnote{note that the
representation space (Fock space) in this case is three
dimensional}:
\begin{equation}
|\xi \rangle_3 = \alpha_0 |0 \rangle + \alpha_1\, \xi \, |1\rangle +
\alpha_2 \, \xi^2 |2\rangle \, ,
\end{equation}
with general coefficients $\alpha_0$, $\alpha_1$ and $\alpha_2$.
Then imposing (\ref{3eigenstates}), on this state and using
(\ref{axi}) permits to determine the coefficients, and the result
is:
\begin{equation}
|\xi\rangle_3 = |0\rangle + j^2 \xi \, |1\rangle + \rho_2^{-1/2} \,
\xi^2\, |2\rangle \; . \label{3CS}
\end{equation}

It is important that, (despite the differences in the definition of
the Grassmann variables and in the relations (\ref{atheta}) and
(\ref{axi})) if we replace $\theta$ by $\xi$ in (\ref{kCS2}) and
$k=3$, (\ref{kCS2}) reduces to equation (\ref{3CS}).

The similarity can be pushed further, as one can rewrite the GGCS
(\ref{3CS}) in the form:
\begin{eqnarray}
|\xi\rangle_3 &=& |0\rangle + a^+ \xi \; |0\rangle + \rho_2^{-1} \;
a^+ \xi a^+\xi \; |0\rangle \nonumber \\&:=& \hbox{exp}_j(a^+\xi) \;
|0 \rangle
\end{eqnarray}
where the generalized exponential function is the same as the one in
(\ref{qexponential}) with $k=3$ or $q=j$.

To conclude the construction of the GGCS in this case, one should
construct a resolution of unity in terms of these states.

Here again one follows the same method as in the previous
subsection. That is,we look for a resolution of unity in the form:
\begin{equation}
\dint d\bar\xi d\xi \; \omega(\bar\xi \xi) \; |\xi \rangle_3 \3bras
\xi| = I\; ; \label{3resolution}
\end{equation}
where $\displaystyle \omega(\bar\xi \xi) = c_0 + c_1 \, \bar\xi \xi
+ c_2 \, \bar\xi \xi \bar\xi \xi$. Then One has to determine the
coefficients $c_0$, $c_1$ and $c_2$ such that the equality
(\ref{3resolution}) holds.

We use the form (\ref{3CS}) of the GGCS, the integration rules and
the relations (\ref{barxi}) and (\ref{axi}) and the completness of
the Fock basis (\ref{qFock}) with $k=3$ in this case:
\begin{equation}
|0\rangle\langle0| + |1\rangle\langle1| + |2\rangle\langle2| = I \;
.
\end{equation}

And the result is the following:
\begin{equation}
\omega(\xi\bar\xi) = -q + \bar\xi \xi + \bar\xi \xi \bar\xi \xi \; .
\end{equation}

\section{Discussions}

Generalized coherent states associated with generalized harmonic
oscillators are defined as eigenstates of the corresponding
annihilation operator. But unlike the Bosonic deformations, where
the representation space is infinite dimensional, the
representation space of the fermionic deformations is finite,
expressing thus, a generalized Pauli exclusion principle. Because
of this fact ordinary variables are not suitable for the
construction of the generalized coherent states. Instead of this
one has to use Grassmann variables to take account of this fact.

We have shown in this paper how to construct GGCS associated to a
generalized deformation of the fermionic oscillator by using the
two generalizations existing in the literature of Grassmann
Grassmann variables: Majid and Kerner's generalized Grassmann
variables.

The result that the form of the GGCS constructed is the same:
\begin{equation}
|\tau \rangle = \hbox{exp}_q(a^+\tau) \, |0\rangle
\end{equation}
where $\tau = \theta$ or $\xi$. The form which is also similar to
that of the generalized coherent states associated to the bosonic
deformation of the harmonic oscillator \cite{1}.

It is worth noting that the construction using Kerner's
$Z_3$-graded Grassmann variables was carried out for the
generalized fermionic oscillator (\ref{qalgebra},
\ref{nilpotency}) for the case $k=3$, i.e. $q= \hbox{exp}\left(
{2\pi i \over 3}\right)$. It is obvious that in order to
generalize this construction  to arbitrarily $k$, one has to
define $Z_k$-graded Grassmann variables in the spirit of kerner's
definition. This is done by imposong a $Z_k$-symmetry over the
generalized Grassmann algebra to be constructed:
\begin{equation}
{\cal{Z}}_k\left( \xi_1 \xi_2 \ldots \xi_k \right) = \xi_1 \xi_2
\ldots \xi_k + \xi_2 \xi_3 \ldots \xi_k \xi_1 +  \ldots + \xi_k
\xi_1 \ldots \xi_{k-1}
\end{equation}
a solution of which is given by the {\it k-nary} relation:
\begin{equation}
\xi_1 \xi_2 \ldots \xi_k = q \; \xi_2 \xi_3 \ldots \xi_k \xi_1
\end{equation}
i.e. under a cyclic permutation of the factors (of a product of
$k$ elements of the algebra) a factor $q$ ($k^th$ root of unity)
appears.

From this relation it follows that:
\begin{itemize}
\item $\displaystyle \left( \xi_i\right)^k= 0$

\item $\displaystyle  \xi_1\xi_2 \ldots \xi_k\xi_{k+1} =
0$
\end{itemize}

One also introduces $\bar\xi$'s duals to the $\xi$'s as
$k-1$-grade elements obeying similar relations as the $xi$'s but
with $q$being replaced by $\bar q = q^{k-1}$. Furthermore the
$\bar \xi$'s obey binary relations with the $^\xi$'s:
\begin{equation}
\xi_i \bar\xi_j = q \; \bar\xi_j \xi_i \; .
\end{equation}
Using then the same relations as in (\ref{axi}) (in fact a similar
reasoning leads to interpret $a^+$ as $k-1$ grade element and $a$
as a 1-grade element).

One can check that the states:
\begin{equation}
|\xi\rangle_k = \hbox{exp}_q(a^+\xi) |0\rangle
\end{equation}
are in fact eigenstates of the annihilation operator with $\xi$ as
an eigenvalue.

\vspace{0.5cm}

It is important to note that the GGCS constructed in this paper, are
very general as on one hand they are related to a generalized
deformation of the fermionic oscillator and it is achieved for any
degree of nilpotency of the operators (and Grassmann variables
used). As a matter of fact these results generalize the results
obtained in \cite{2} and \cite{3}.

\end{document}